\pdfoutput=1
\documentclass{article}

\usepackage{graphicx} 
\usepackage{bmpsize}
\usepackage{amsmath,amssymb}
\begin{document}
\title{Sketched Floor plans versus SLAM maps: A Comparison}
\author{Leo Bowen-Biggs\thanks{Worchester Polytechnic Institute} \and Suzanne Dazo\thanks{University of Nebraska}
Yili Zhang\thanks{Oregon State University} \and Alex Hubers\thanks{Cornell College} \and Matthew Rueben\thanks{Oregon State University} \and Ross Sowell\thanks{Cornell College} \and William D. Smart\thanks{Oregon State University} \and
Cindy Grimm\thanks{Oregon State University}}

\maketitle              % typeset the title of the contribution

\begin{abstract}
Maps --- specifically floor plans --- are useful for a variety of tasks from arranging furniture to designating conceptual or functional spaces (e.g., kitchen, walkway). We present a simple algorithm for quickly laying a floor plan (or other conceptual map) onto a SLAM map, creating a one-to-one mapping between them. Our goal was to enable using a floor plan (or other hand-drawn or annotated map) in robotic applications instead of the typical SLAM map created by the robot. We look at two use cases, specifying ``no-go'' regions within a room and locating objects within a scanned room. Although a user study showed no statistical difference between the two types of maps in terms of performance on this spatial memory task, we argue that floor plans are closer to the mental maps people would naturally draw to characterize spaces.
\end{abstract}
\section{Introduction}
SLAM maps are ubiquitous in robotic applications, in part because they are relatively simple to make and useful (from a robot point of view) for localization and navigation. However, we argue that SLAM maps are not ``natural'' for most tasks such as labeling a table location or selecting a desired pathway for the robot to use~\cite{Shah01112013,boniardi2016icra}. 

\begin{figure}
\begin{center}
\includegraphics[width=0.95\linewidth,natwidth=488,natheight=240]{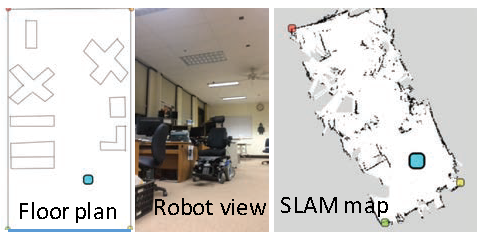}
\caption{Left: A hand-drawn floor plan of the space, robot location shown by blue rectangle. Middle: Robot's first person point of view. Right: SLAM map created by driving the robot around the space.}
\label{fig:SLAMtoNice}
\end{center}
\end{figure}

SLAM maps have two properties that make them unnatural for humans. First, SLAM maps have a lot of noise and spurious points even along relatively simple walls. Second, SLAM maps often have global distortion, creating, eg, a C-shaped room out of a long narrow one (see Figure~\ref{fig:SLAMtoNice}). This forces the viewer to mentally ``undo'' the local rotation. Existing research shows that extra visual information on traditional street maps adversely affects performance on a navigation task~\cite{Sanchez2009}. We informally validated that, even with a SLAM map available as an example, people drew floor plans that were similar to the floor plan shown on the left in Figure~\ref{fig:SLAMtoNice}.

With this in mind we describe a simple algorithm for establishing a correspondence between a sketched floor plan (or real one) and a SLAM map. We demonstrate two use cases: Mapping navigation information to the floor plan and mapping ``no-go'' regions from the floor plan to the SLAM map to prevent the robot from entering a designated area.

We conducted a user study looking at the effectiveness of the floor plan over a SLAM map for a spatial memory task. There was no statistically significant difference in the average performance, however there was a slight difference in the kinds of errors the participants made.

%%%%%%%%%%%%%%%%%%
%%%%%%%%%%%%%%%%%%

\section{Background}
%[check for someone who’s done something to map 3D geometry to slam map <-- what does this mean, Cindy?]
% I was assuming someone, somewhere, has had to map a floor plan to a slam map; not sure how they do that

% SLAM maps, defined
Many robots use a 2D, metric map to navigate, where obstacles and free spaces are represented as an occupancy grid~\cite{elfes_using_1989}. We call these \emph{SLAM maps} after the simultaneous localization and mapping (SLAM) algorithms commonly used to create them (see, e.g.,  Dissanayake et al~\cite{dissanayake_solution_2001} for early formulation and solution of the SLAM problem). 

%(NOTE: other statements about how SLAM maps aren't great for humans are in the introduction.)

% Floor plans, defined
We compare SLAM maps with sketched \emph{floor plans}, which are similar to building blueprints and the more simplified maps found, e.g., on shopping mall directories. We consider floor plans to be semi-accurate: the walls are (relatively) straight on the map when they are straight in real life, and relative sizes and angles are all about correct (although aspect ratio and overall scales may be wrong). The level of detail is typically somewhat low, specialized for navigation: mostly just the walls, doors, and major obstacles are shown. Floor plans also lack the sensor errors that cause holes in walls and false obstacles in the middle of a room in SLAM maps. Significantly, floor plans are typically drawn with lines and curves, not by filling the cells of a grid. 

% For the 1st study (SLAM vs. semantic) ... at least the part about reading maps. 
Pinheiro~\cite{pinheiro_determinants_1996} includes a review of how people read and draw maps. Among the concepts reviewed is ``naive cartographic realism'', which is when map readers assume that the map perfectly represents the real world. 

% For the 2nd study: "watch a video, sketch a map"
%  [NOTE: I can only access the abstract of this paper. Cindy?]
For millenia people have been creating maps by simply walking through a space to construct a \emph{mental map} of it, which is later transferred to paper as a \emph{sketch map}. Beck and Wood~\cite{beck_cognitive_1976} include several common operations involved in map-making in their model of urban mapping. These include scaling, rotation, and synchronization of observations from several tours of a space. Pinheiro~\cite{pinheiro_determinants_1996} notes that people tend to create maps hierarchically by first subdividing the region to be mapped; this is much like we divide the visible stars into constellations, then map their relative locations on a per-constellation basis. Kuipers~\cite{kuipers_map_1982} asks whether people encode spatial information as a ``Map in the Head'' such that your sketch map is just like your mental map. It seems this is not entirely the case: mental maps can have sections that are disconnected, and routes could be represented such that they are only valid in one direction. Billinghurst and Weghorst~\cite{billinghurst_use_1995} test whether sketch maps are valid measures of certain aspects of mental (``cognitive'') maps. They found that the accuracy of a sketch map is highly correlated with world knowledge and the subjective feeling of orientation in a space. Similarly, Wang and Li~\cite{WangLiSketch2012} found that sketched maps were more accurate than verbal instructions for navigating from one point to another.

Recently, researchers have been using sketched maps directly in localization tasks without an intermediate SLAM representation~\cite{boniardi2016icra,Shah01112013,SetalaphrukSketchMap2003}. This is relevant when, for example, it is not feasible to build a SLAM map first. From an interface standpoint, it may make sense to ask the user to use our technique to establish a rough correspondence between any (incomplete) SLAM map the robot creates during its localization process in order to help guide the robot.

Researchers have also aimed to automate matching of a hand-drawn map to a SLAM or other automatically generated map for the purposes of navigation~\cite{Boniardi2015,MatsuoOutdoorMap2012,SpringBasedMatching2007}. In the first case, the robot simultaneously locates itself on both the SLAM map and the sketch, including a scale factor for the sketch map to account for inaccuracies in the sketch. In the second case the sketch is used primarily to seed the localization procedure with estimated building locations. In the third case they assume a one-to-one mapping between the objects in the map and the objects in the sketch (i.e., object correspondence, not spatial) and that the sketch and real map have the same number of objects. We sidestep this problem by asking the user to establish correspondences for us. Inaccuracies in the sketch map are essentially represented as local affine transformations (given by the distortions in the triangle shapes from one map to the other).

\section{User Interface and Algorithm}
We describe the interface and algorithm and its implementation.

\begin{figure}
\begin{center}
\includegraphics[width=0.95\linewidth,natwidth=1002,natheight=529]{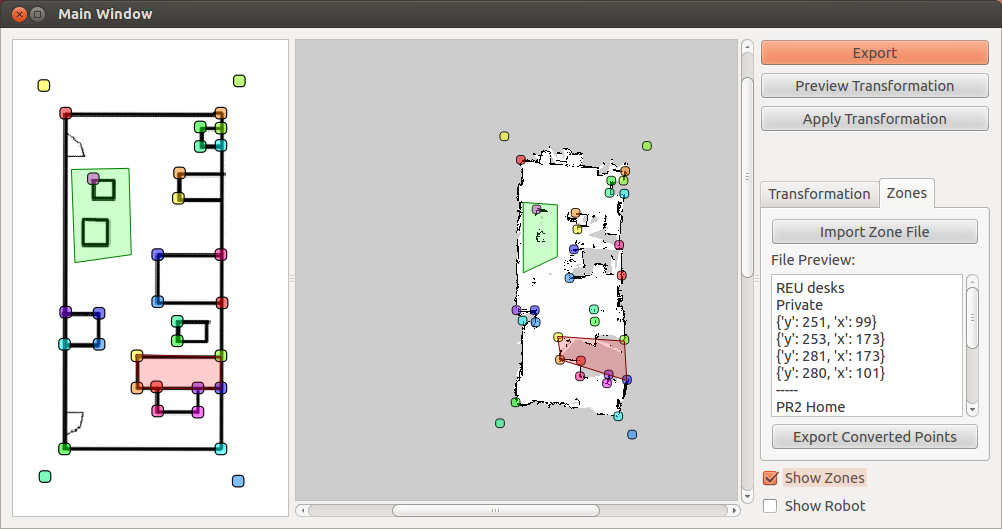}
\caption{Establishing correspondences between a floor plan and the SLAM map (colored squares).  The red and green regions are the spaces marked on the floor plan that are mapped (automatically) back to the SLAM map.}
\label{fig:NoGo}
\end{center}
\end{figure}

From the user' point of view they simply click corresponding points in the two maps, approximately one for every corner of the floor plan (see Figure~\ref{fig:NoGo} and four corners for Figure~\ref{fig:SLAMtoNice}). The system then calculates a one-to-one and onto mapping between the two maps. After this the user can mark a point or line in one map and have it appear in the other one.

The interface and algorithm were implemented in the Robot Operating System (ROS). We demonstrate two use cases. The first case maps the robot's position and orientation from the SLAM map to the floor plan during a robot navigation task (see Figure~\ref{fig:userStudy} and accompanying video). In the second use case we used the floor plan to mark a ``no go'' region, then mapped this back to the SLAM map as walls. The robot then drove around the marked region (see Figure~\ref{fig:NoGo} and accompanying video).

%%%%%%%%%%%%%%%%%%
%%%%%%%%%%%%%%%%%%

\subsection{Algorithm}
Essentially, we use Triangle~\cite{Triangle} to triangulate the floor plan then map the triangle vertices to the SLAM map using the user-marked coordinates. To calculate the correspondence we use barycentric coordinates within each triangle. Although this introduces discontinuities in the derivatives along the boundaries of the triangles we have not found this to be a problem in practice, probably because the local deformation from one triangle to the next is fairly minimal (and the SLAM map is noisy, hiding small inaccuracies). If desired, the user can draw two curves which will be automatically split into a poly line when the poly line varies too much from either curve; matching is accomplished using arc-length parameterization.

\section{User Studies and Evaluation}
We discuss two methods for validation; support for floor plans or conceptual maps, and a comparison of the functionality of a floor plan versus a SLAM map for spatial localization.

\subsection{Support for floor plans or conceptual maps}

\begin{figure*}
\begin{center}
\includegraphics[width=0.95\linewidth,natwidth=6027,natheight=1871]{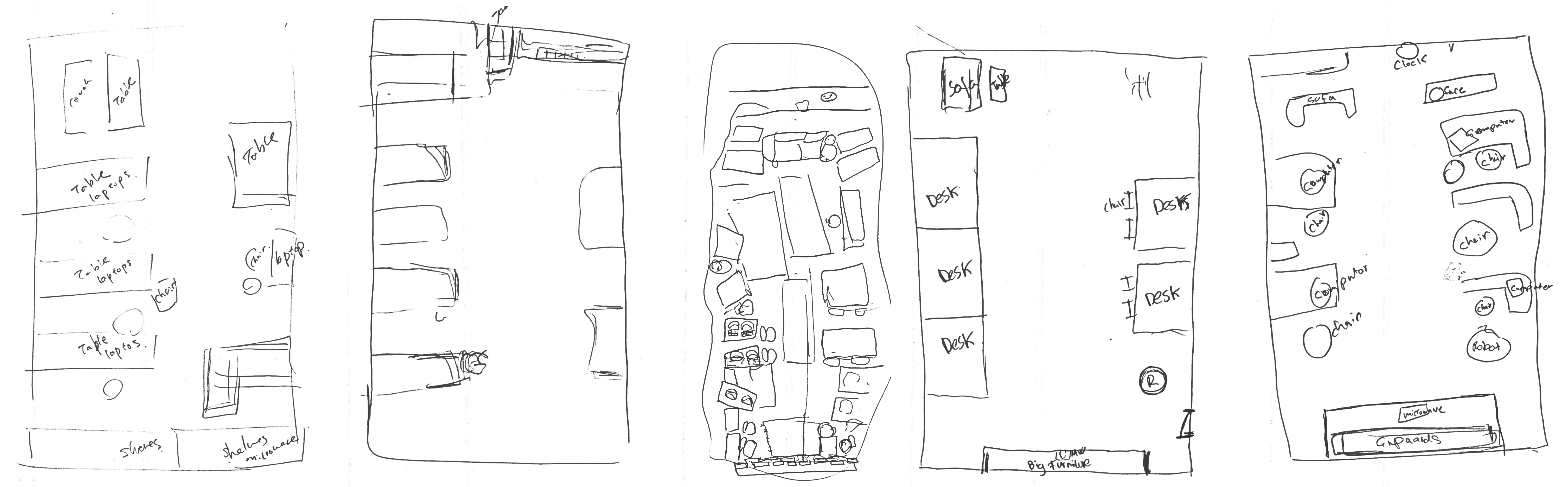}
\caption{Five hand-drawn maps of the space shown in the first figure. Images have been darkened after scanning (they were pencil drawings).}
\label{fig:handDrawn}
\end{center}
\end{figure*}

Clearly it is possible to use a SLAM map, however, we argue that it does not match people's conceptual models of a space. To provide evidence for this we asked five non-roboticists to draw a map of a space the robot had mapped. The participants were given a video from the view point of the robot as it drove through a previously mapped space. The participants saw the robot's location in the SLAM map at the same time (middle and right image of Figure~\ref{fig:SLAMtoNice}). Participants were told they would see a video of the map, with a robot-created map on the side showing where the robot was as a blue square. After watching the video as often as you need to do, please draw the outline of the room and the locations of all the objects in the room, as best as you remember.  The five maps are shown in Figure~\ref{fig:handDrawn}). 

%%%%%%%%%%%%%%%%%%
%%%%%%%%%%%%%%%%%%

%From Yili: I told every map drawer that I would show you a video of a room, on the left side is a slam map and is on the top view of the room; the blue square on the map represents the location of the robot. On the right side is a video of the room, which is from the perspective of the robot (blue square). The video is gonna take 1 min and 30 seconds, please watch the video carefully and after watching I would ask you to draw the out line of the room and the locations of all the objects in the room as specific as you can based on your memories from the video.

%%%%%%%%%%%%%%%%%%
\begin{figure*}
\begin{center}
\includegraphics[width=0.95\linewidth,natwidth=1004,natheight=613]{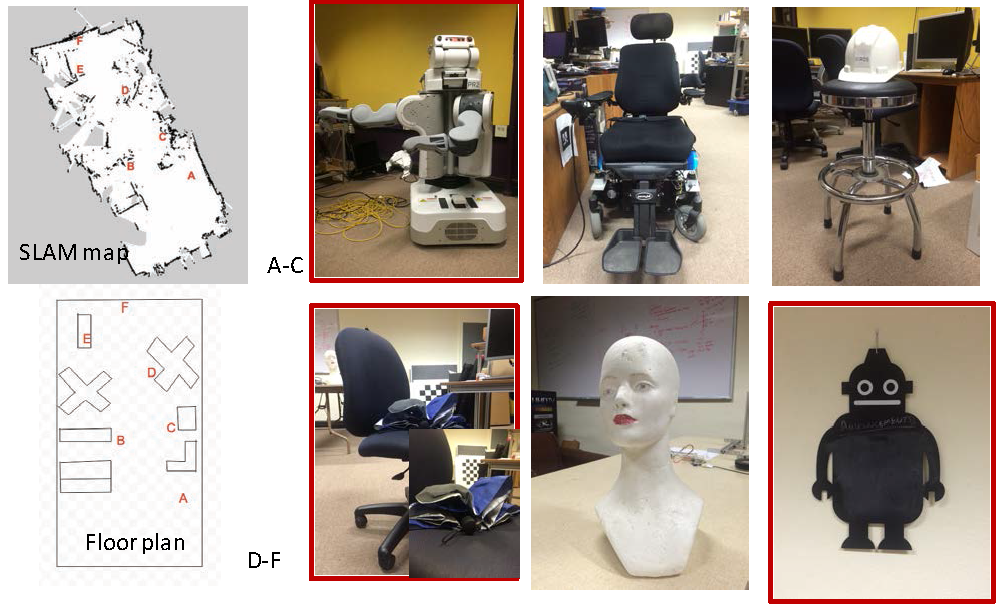}
\caption{Left: The SLAM and floor plan with the letter locations marked. Objects are shown in the order they are encountered (A-F). Images with a red boundary were shown along with the question; the actual video frame for the hat and umbrella are also shown. Accuracy in order: 60\%, 61\%, 37\%, 53\%, 51\%,	61\%.}
\label{fig:userStudy}
\end{center}
\end{figure*}
%%%%%%%%%%%%%%%%%%

Although the hand-drawn maps are all different, there are some commonalities. Everyone included the tables and the shelves in the back of the room (although the number of tables varies). Note that the tables are not really identifiable in the SLAM map. Interestingly, one person (second drawing) flipped the map so that the starting point of the robot's path was at the top. Three participants included the sofa. It appears one person attempted to preserve the curvi-linear boundary of the SLAM map; this map also had the most detail. Despite being asked to include objects, participants primarily included furniture (chairs, tables, sofas), with the only objects being marked computers and robot. The aspect ratio of the space also varies, with only the most detailed map having a similar aspect ratio to the SLAM map (other maps are not as narrow).

Detailed analysis of how people conceptualize space is beyond the scope of this paper, however, our informal study supports the idea that people visualize space using floor plan-like layouts.

\subsection{Spatial mapping user study}
\label{sec:SMUserStudy}

We next describe our spatial mapping on-line user study.

\subsubsection{Study stimulus and design} 
\label{sec:studyDesign}

We made two videos, one with the SLAM map and one with the floor plan (see Figure~\ref{fig:SLAMtoNice}). In each video the participant saw the map on the left and the robot's point of view on the right. Participants were instructed to ``Please watch this video carefully and pay attention to what you see where (you may watch as many times as you want)''.  They could watch the video as many times as they wanted, but could not go back to the video once they started answering questions. Video length was one and a half minutes, and consisted of a navigation from one end of the room to the other and back, avoiding the obstacles in the room and pausing to look at the objects in the study.

We asked one open-ended question (name three objects in the video) to ensure that participants had actually watched the video. There were four questions about objects in the room and six questions that asked the participants to identify the location of an object in the video. For potentially unclear objects (the robot picture, umbrella, and PR2) a picture of the object was included with the text question. Participants picked the location from one of six on the map (letters A-F). Each participant saw all of the questions and all of the objects, randomly ordered.

The four questions were, with the correct answer and the number of people answering that question correctly:

\begin{itemize}
\item How many actual robots did you see (not pictures of robots)? (Answer: 1, 82\%)
\item How many clocks did you see?  (Answer: 1, 56\%)
\item  What is the color of the couch? (yellow, brown, dark, pink) (Answer: brown, confounded with dark, 88\%)
\item Where is the whiteboard? (on the wall, beside the door, in the middle of the room) (Answer: on the wall, 65\%)
\end{itemize}

The six objects are shown in Figure~\ref{fig:userStudy}, along with the SLAM and floor plan with letter locations labeled.

We pilot tested the study with 7 people, asking them to talk aloud while taking the study. We used this to ensure that the questions were both clear and answerable the majority of the time. We explicitly checked that questions were answered correctly some of the time, but not always (i.e., they were of mid-level difficulty).

\subsubsection{Participants}

We ran the study on-line using Amazon's Mechanical Turk (70 participants), of which 70 attempted the task (30 floor plan, 34 SLAM)). Of those, 24 successfully completed the floor plan condition (80\%), 31 the SLAM map (91\%). Successfully means they answered all of the questions. The remainder quit after watching the video or answered, at most, one or two questions.

\subsubsection{Results of on-line study}

%%%%%%%%%%%%%%%%%%
\begin{figure}
\begin{center}
\includegraphics[width=0.95\linewidth,natwidth=946,natheight=767]{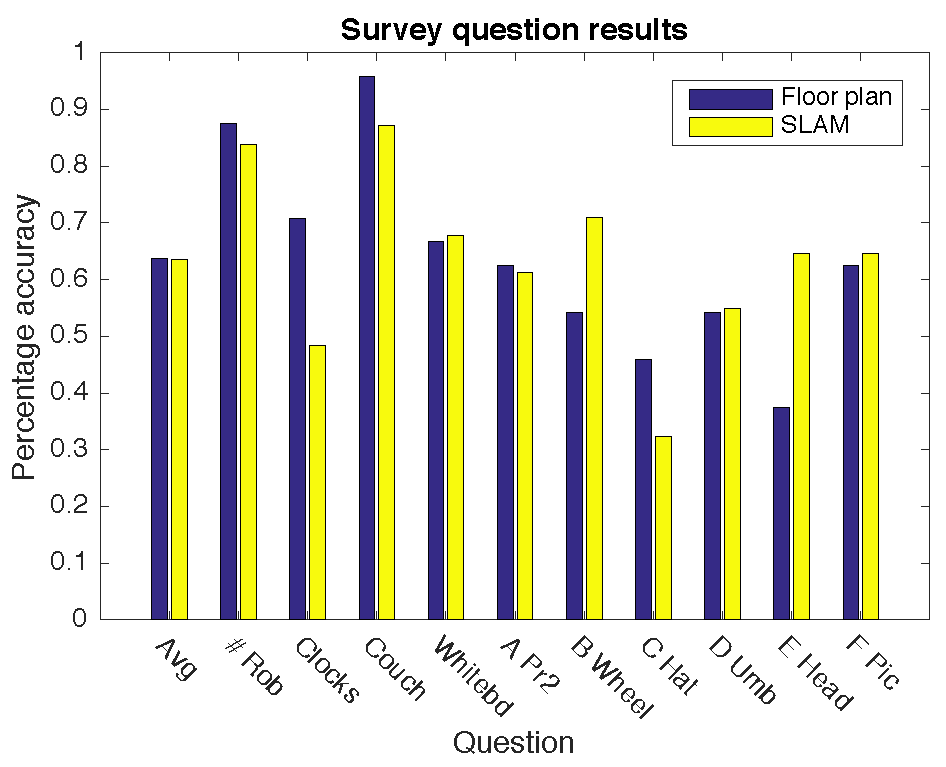}
\caption{Percentage of participants who correctly answered the questions and located the objects on the map for each condition (average of percentages on the left).}
\label{fig:surveySummary}
\end{center}
\end{figure}
%%%%%%%%%%%%%%%%%%

\begin{figure*}
\begin{center}
\includegraphics[width=0.95\linewidth,natwidth=1639,natheight=862]{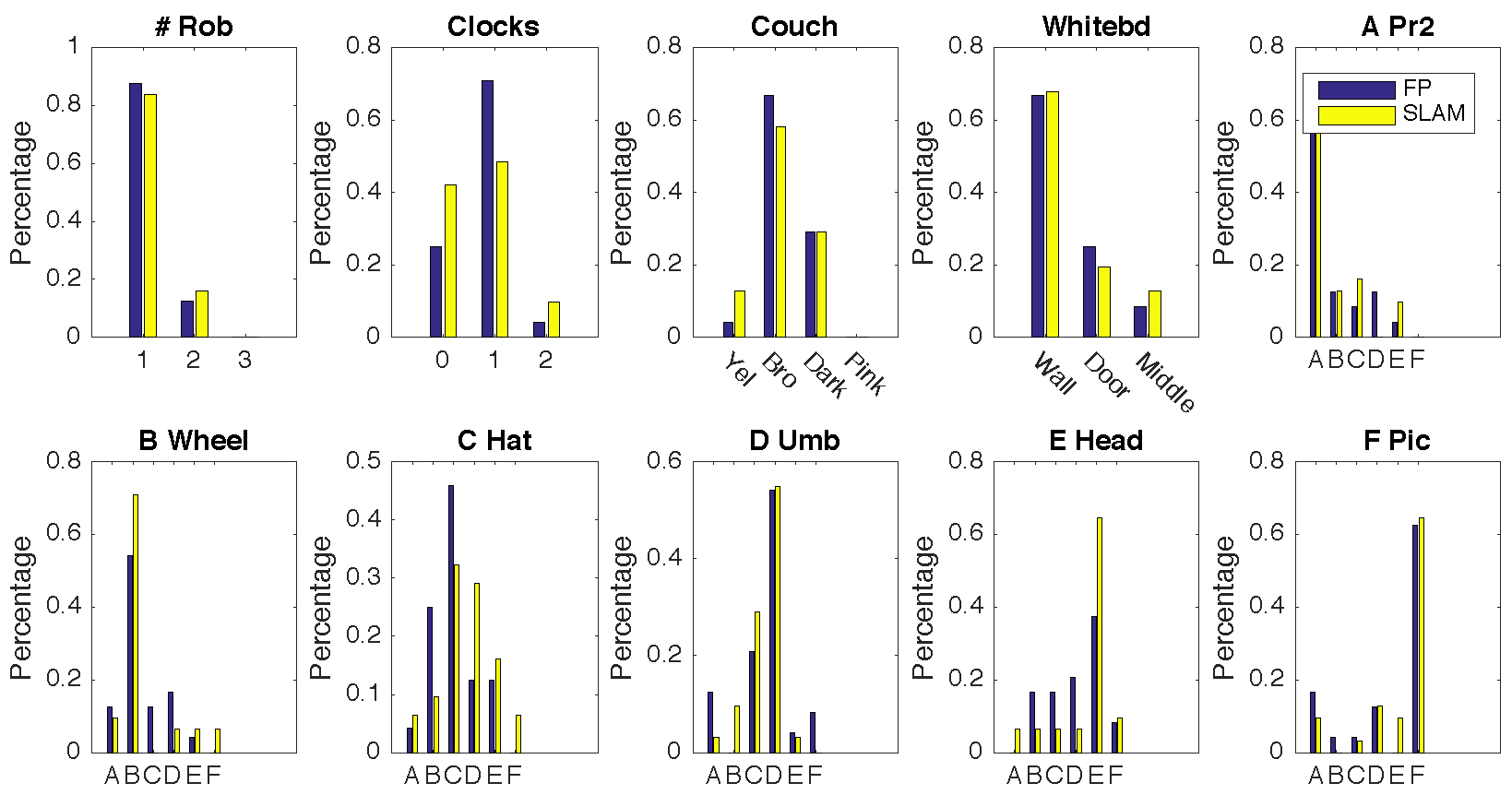}
\caption{Distribution of answers per question, given as percentages.}
\label{fig:surveyQuest}
\end{center}
\end{figure*}

We summarize the mean percentage correct for all questions in Figure~\ref{fig:surveySummary}; note that these are all essentially correct/incorrect questions, so there is no standard deviation. Although the mean for the floor plan percent correct is very slightly better --- and better for the four questions -- the results are not statistically significant. There was some variation in {\em how} participants got the answers wrong (see Figure~\ref{fig:surveyQuest}. In general the answer distribution is roughly bell shaped (recall that the A-F places the objects roughly in order from bottom to top) which implies that participants were usually close. 15 of the participants used all six locations exactly once; half of these were correct (split equally across both conditions). No one got all questions correct.

%%%%%%%%%%%%%%%%%%
%%%%%%%%%%%%%%%%%%

\section{Discussion}
The literature (and our own study) clearly shows that, when asked to provide a sketch of the layout of a space people tend to provide simple line drawings. So why did the on-line study show no {\em functional} difference in a spatial memory location task? We hypothesize that this task is difficult enough --- and people's abilities to visually build a mental model of a space from a first-person view robotic drive through vary enough --- that any potential benefit of a floor plan is lost in the noise. 

We would argue that most people would prefer a floor plan, and that it makes ``conceptual’’ tasks — such as specifying a path from Bob’s office to the coffee pot — easier. It is not clear that {\em having} a floor plan instead of SLAM map will substantially improve performance for spatial reasoning tasks (although {\em drawing} such a map, possibly with the help of a SLAM map, might). Determining how the artifacts of a SLAM map (global warps, spurious noise) interfere (if they do) with spatial reasoning is beyond the scope of this paper, but a potentially interesting area for future work.

Our current implementation is polygon-based. There are more elaborate 2D mapping/morphing schemes~\cite{Poranne:2014:PGP:2601097.2601123,Alexa:2000:ASI:344779.344859} that could be used instead. These approaches would yield a continuous mapping but are more computationally intensive to evaluate and may not have a well-defined inverse. It would also be possible to apply sketch beautification techniques (either internally or visible to the user) to ``clean up’’ hand-drawn maps~\cite{Cheema:2012:QID:2207676.2208550,Fiser:2015:SDB:2810210.2810215}. This might make it easier for a robot to use the hand-drawn sketch directly for localization.

\section{Conclusion}

We have presented a simple technique for mapping a hand-drawn sketch or floor plan to a SLAM map. This provides a more ``user-friendly’’ experience for labeling SLAM maps and communicating spatial information to the robot. 
%
% ---- Bibliography ----
%
\section*{Acknowledgements}

Funding agencies and friends.

\bibliographystyle{alphanc}
\bibliography{visualsMatter,mapBackground,mendelay}
\end{document}